\documentstyle[12pt]{article}
\makeatletter \@addtoreset{equation}{section}

\makeatother
\begin{document}
\begin{center}
{\bf Kepler Problem in the Constant Curvature Space.} \\

{\bf G.P.Pronko}\\

{\it Institute for High Energy Physics , Protvino, Moscow
reg.,Russia}\\
{\it Institute of Nuclear Physics, National Research Center
"Demokritos", Athens, Greece}

\end{center}

\begin{abstract}
 We present algebraic derivation of the result of Schr\"{o}dinger \cite{Schrod} for the
 spectrum of hydrogen atom in the space with constant curvature.
\end{abstract}

\section{Introduction}

The theory of different mechanical systems, both classical and
quantum in the space of constant curvature attracts attention since
long. Especially interesting to study the systems like Hydrogen
atom, initiated by Schr\"{o}dinger \cite{Schrod} in this curved
space, as it may have cosmological application. A collection of
references concerning these issues could be found in
\cite{Santander}. Here we are not going to discuss neither
motivation, no application of this system. What we are going to
consider is the dynamical symmetry of Hydrogen atom in the space
with constant curvature in classical and especially in quantum case
and, as it is possible to do in the flat case \cite{Fock} to derive
the spectrum of the system by pure algebraic tools. It happened to
be quite unexpected that in spite of similarity of two system --- in
flat and curved spaces, the construction of its generators in the
curved case is not at all straightforward.

The 3-dimensional space with constant curvature are sphere $S^3$,
hyperboloids  $H^3,H'^3$ and cone $V^3$. For definiteness we shall
consider the case of positive curvature e.d. $S^3$. The coordinates
of $S^3$ will be $X_{\alpha}$, $\alpha=1,2,3,4$, $X_{\alpha}^2=1$.
The simplest mapping of $R^3$ ($R^3/\{\infty\}$) into $S^3$ is
stereographic projection given by
\begin{equation}\label{1}
X_{\alpha}=\left( \frac{2\lambda {\bf x}}{{\bf x^2}+\lambda^2},\quad
\frac{{\bf x^2}-\lambda^2}{{\bf x^2}+\lambda^2}\right)
\end{equation}
The only possible kinetic term, invariant with respect to group of
motion of $S^3$---$SO(4)$ arises if we take the Lagrangian
proportional to the square of angular momentum tensor $M_{\alpha
\beta}=X_{\alpha}{\dot X}_{\beta}-{\dot X}_{\alpha}X_{\beta}$:
\begin{equation}
L=\frac{m}{2}\frac{1}{4\lambda^2}M_{\alpha\beta}^2=\frac{m}{2}\frac{{\bf\dot
x^2}}{({\bf x^2}+\lambda^2)^2}
\end{equation}
The kinetic term of the Hamiltonian, which corresponds to this
Lagrangian is given by
\begin{equation}
H=\frac{1}{2m}{\bf p^2}({\bf x^2}+\lambda^2)^2
\end{equation}
This Hamiltonian has 6 integrals of motion, which are generators of
its group of motion $SO(4)$:
\begin{eqnarray}
L_{i}&=&\epsilon_{ijk}x_j p_k,\nonumber \\
K_{i}&=&\frac{1}{2\lambda}(p_i({\bf x^2}-\lambda^2)-2 x_i {\bf
px}),\nonumber
\end{eqnarray}
The canonical Poisson brackets for $p_i,x_i $ induce the algebra
$SO(4)$ for generators $L_i,K_i$:
\begin{eqnarray}
&\{L_i,L_j\}=-\epsilon_{ijk}L_k\nonumber\\
&\{K_i,K_j\}=-\epsilon_{ijk}L_k\nonumber\\
&\{L_i,K_j\}=-\epsilon_{ijk}K_k\nonumber
\end{eqnarray}
The Hamiltonian $H$ proportional to one Casimir:
\begin{equation}
H=\frac{1}{2m}4\lambda^2(\bf L^2+K^2),\nonumber
\end{equation}
while the second $\bf LK=0$.

Now we shall consider Hamiltonian with Kepler potential
\cite{Schrod},\cite{Santander}
\begin{equation}\label{5}
H=\frac{1}{2m}{\bf p^2}({\bf
x^2}+\lambda^2)^2+\frac{2\lambda\alpha}{m}\frac{{\bf
x^2}-\lambda^2}{|{\bf x}|}
\end{equation}
As in flat case, the Hamiltonian (\ref{5}) possesses 6 integrals of
motion (of course they are not all independent): the angular
momentum $\bf L$ and Laplace-Runge-Lentz (LRL) vector $\bf A$
\begin{equation}
\bf A=K\times L+\alpha \frac{x}{|x|},
\end{equation}
apparently $\bf A$ and $\bf L$ satisfy the relation
\begin{equation}
\bf A L=0
\end{equation}
The algebra of Poisson brackets of vectors $\bf A, L$ has the
following form:
\begin{equation}\label{8}
\{A_i,A_j\}=\epsilon_{ijk}L_k(\frac{m}{2\lambda^2}H-2{\bf
L^2}),\qquad \{L_i,A_j\}=-\epsilon_{ijk}A_k.
\end{equation}
The square of vector $\bf A$ is expressed via $\bf L^2$ and
Hamiltonian:
\begin{equation}\label{9}
{\bf A^2}=\alpha^2-{\bf L^4}+\frac{m}{2\lambda^2}{\bf L^2}H.
\end{equation}
The reader, acquainted with properties of LRL in the flat case
\cite{Fock} immediately sees the difference both in algebra and the
relation of $\bf A^2$ with $H$ and $\bf L^2$. This difference does
not allow us to define in a simple way the vector which forms
together with $\bf L$ the $SO(3)\times SO(3)$ algebra which is
dynamical symmetry responsible for supplementary degeneration in
quantum mechanics. But this construction is however possible and
consideration of its classical analog shows the way towards it in
quantum case. First of all let us introduce the rescaled Hamiltonian
$h$:
\begin{equation}
h=\frac{m}{2\lambda^2}H.
\end{equation}
From equation (\ref{9}) we obtain
\begin{equation}\label{11}
{\bf A^2}=-[{\bf
L^2}-(\frac{h}{2}+\sqrt{\frac{h^2}{4}+\alpha^2})][{\bf
L^2}-(\frac{h}{2}-\sqrt{\frac{h^2}{4}+\alpha^2})]
\end{equation}
and, because ${\bf A^2}>0$, $\bf L^2$ satisfies the following
inequality:
\begin{equation}\label{12}
0\leq {\bf L^2}\leq (\frac{h}{2}+\sqrt{\frac{h^2}{4}+\alpha^2}).
\end{equation}
Now let us find the Poisson brackets of ${\bf A}f({\bf L^2})$ (the
function $f({\bf L^2})$ could be $h$-dependent)
\begin{equation}\label{13}
\{A_if({\bf L^2}),A_jf({\bf L^2})\}=\epsilon_{ijk}L_k \frac{\partial
({\bf A^2}f^2({\bf L^2}))}{\partial{\bf L^2}}
\end{equation}
In order for (\ref{13}) be the part of $SO(3)\times SO(3)$ algebra
the following condition should be satisfied
\begin{equation}\label{14}
{\bf A^2}f^2({\bf L^2})=a-\bf L^2.
\end{equation}
Apparently, if we take
$a=(\frac{h}{2}+\sqrt{\frac{h^2}{4}+\alpha^2})$ the R.H.S of
(\ref{14}) will be positive for any point in the phase space because
of inequality (\ref{12}). Moreover, the function $f({\bf L^2})$ for
this choice  of $a$ will be essentially simplified
\begin{equation}\label{15}
f({\bf L^2})=\sqrt{\frac{a-{\bf L^2}}{{\bf A^2}}} =\left[{\bf
L^2}-(\frac{h}{2}-\sqrt{\frac{h^2}{4}+\alpha^2})\right]^{-\frac{1}{2}}
\end{equation}
Thus, the vector
\begin{equation}
{\bf R}={\bf A}\left[{\bf
L^2}-(\frac{h}{2}-\sqrt{\frac{h^2}{4}+\alpha^2})\right]^{-\frac{1}{2}}
\end{equation}
together with angular momentum $\bf L$ forms the desired algebra
$SO(3)\times SO(3)$:
\begin{eqnarray}
&\{L_i,L_j\}=-\epsilon_{ijk}L_k,\nonumber\\
&\{R_i,R_j\}=-\epsilon_{ijk}L_k,\nonumber\\
&\{L_i,R_j\}=-\epsilon_{ijk}R_k.
\end{eqnarray}
Further, the following expression  for nontrivial Casimir of
$SO(3)\times SO(3)$  holds true:
\begin{equation}\label{18}
{\bf R^2+L^2}=(\frac{h}{2}+\sqrt{\frac{h^2}{4}+\alpha^2}),
\end{equation}
the other Casimir ${\bf RL}$ vanishes. From (\ref{18})we obtain
\begin{equation}
h={\bf R^2+L^2}-\frac{\alpha^2}{{\bf R^2+L^2}}.
\end{equation}
This is the result we were looking for and which is of a great
importance in quantum case. The obvious obstacle for quantum
generalization of this construction is non-commutativity of
operators $\bf A$ and $\bf L^2$. This will be the problem we are
going to solve below.

\section{ Quantum Theory}

The Hilbert space of quantum theory of Kepler problem in the space
of constant curvature $S^3$ is the space \quad$L^2$ on $S^3$, where
the scalar product $<\phi|\psi> $ is given by
\begin{equation}\label{20}
<\phi|\psi> =\int d^4X \bar\phi(X)\psi(X)\delta(X^2-1),
\end{equation}
where $X$ are coordinates in 4D space $X_{\alpha},\alpha=1,2,3,4$.
Using the map $R^3$  into $S^3$ given by (\ref{1}) we can express
the scalar product (\ref{20}) up to inessential factor as an
integral over $R^3$:
\begin{equation}\label{21}
<\phi|\psi>=\int d^3x \frac{1}{({\bf x^2}+\lambda^2)^3}\bar\phi({\bf
x})\psi({\bf x})
\end{equation}
The measure in this scalar product makes non-trivial definition of
operators we need for Kepler problem. Apart from making these
operators Hermitian we have to preserve their  algebraic properties.
As an example let us consider the vector $\bf K$. If we define the
quantum operator as
\begin{equation}\label{22}
 K_i=\frac{1}{2\lambda}\left[({\bf x^2}-\lambda^2)p_i-2x_i\bf xp\right],
\end{equation}
where the operator ${\bf p}=-i\frac{\partial}{\partial {\bf x}}$
(note, that $\bf p$ is not Hermitian), then $\bf K$ will be
Hermitian and its commutation relations will be the same, as in
classical case:
\begin{equation}
[K_i,K_j]=i\epsilon_{ijk}L_k
\end{equation}
The operator $\bf L$ brings no difficulty because it commutes with
measure in (\ref{21}). Having both operators $\bf K,L$ properly
defined we can find the kinetic part of quantum Hamiltonian:
\begin{equation}
{\bf K^2+L^2}=-\frac{({\bf
x^2}+\lambda^2)^3}{4\lambda^2}\frac{\partial}{\partial
x_i}\frac{1}{({\bf x^2}+\lambda^2)}\frac{\partial}{\partial x_i},
\end{equation}
which is apparently Hermitian. Thus, the total Hamiltonian has the
following form:
\begin{equation}
H=\frac{2\lambda^2}{m}\left[{\bf
K^2+L^2}+\frac{\alpha}{\lambda}\frac{{\bf x^2}-\lambda^2}{|\bf x|}
\right]=\frac{2\lambda^2}{m}h
\end{equation}
Now we are ready to define the quantum LRL vector
\begin{equation}
{\bf A}=\frac{1}{2}\left(\bf K\times L-L\times
K\right)+\alpha\frac{\bf x}{|\bf x|}
\end{equation}
With this ordering the operator $\bf A$ commutes with Hamiltonian:
\begin{equation}
[h,\bf A]=0
\end{equation}
The commutator of different components of $\bf A$ are the same as in
classical case:
\begin{equation}
[A_i,A_j]=-i\epsilon_{ijk}L_k(h-2{\bf L^2}),
\end{equation}
while its square is slightly different
\begin{equation}\label{29}
{\bf A^2}=\alpha^2+h({\bf L^2}+1)-({\bf L^2}+1)^2+1.
\end{equation}
Now we will be looking for the operator $\bf R$, such
that
\begin{equation}\label{30}
[R_i,R_j]=i\epsilon_{ijk}L_k
\end{equation}
in the form, which does not destroy the Hermicity of $\bf R$:
\begin{equation}
R_i=f^{1/2}({\bf L^2})A_if^{1/2}({\bf L^2})
\end{equation}
Further it will be convenient to use instead of operator $\bf L^2$
its function $\gamma$:
\begin{equation}
\gamma=\sqrt{({\bf L^2}+\frac{1}{4})}-\frac{1}{2},\qquad {\bf
L^2}=\gamma(\gamma+1).
\end{equation}
Also we shall need formula, which is proven in the {\bf Appendix},
valid for any vector operator $\bf A$, such that $\bf AL=0$:
\begin{eqnarray}\label{33}
A_if(\gamma)=\frac{f(\gamma+1)}{2\gamma+1}\left[(\gamma+1)A_i+i\epsilon_{ijk}L_jA_k
\right]\nonumber\\
+\frac{f(\gamma-1)}{2\gamma+1}\left[\gamma A_i-i\epsilon_{ijk}L_jA_k
\right]
\end{eqnarray}
Now we are ready to calculate the commutator (\ref{30}):
\begin{eqnarray}\label{34}
\displaystyle&[f^{1/2}(\gamma)A_if^{1/2}(\gamma),f^{1/2}(\gamma)A_jf^{1/2}(\gamma)]&\nonumber\\
&=f^{1/2}(\gamma)\left(A_i f(\gamma)A_j-A_j
f(\gamma)A_i\right)f^{1/2}(\gamma).
\end{eqnarray}
To calculate the parenthesis in the R.H.S. of (\ref{34}) we use the
equation (\ref{33}) and after some algebraic transformation we
arrive at
\begin{eqnarray}\label{35}
&\left(A_i f(\gamma)A_j-A_j f(\gamma)A_i\right)=\nonumber\\
\displaystyle&i\epsilon_{ijk}L_k\left(\frac{f(\gamma+1)}{2\gamma+1}(\gamma
r(\gamma)-{\bf A^2})+\frac{f(\gamma-1)}{2\gamma+1}((\gamma+1)
r(\gamma)+{\bf A^2})\right),
\end{eqnarray}
where we have introduces function $r(\gamma)$ through
\begin{equation}
[A_i,A_j]=i\epsilon_{ijk}L_k r(\gamma)
\end{equation}
Making use of this result we proceed with calculation of commutator
(\ref{30}) and move all $f^{1/2}(\gamma)$ to the right, because they
commute with $\bf L$
\begin{equation}
[R_i,R_j]=i\epsilon_{ijk}L_k T,
\end{equation}
where $T$ is given by
\begin{equation}
T=\frac{f(\gamma)f(\gamma+1)}{2\gamma+1}(\gamma r(\gamma)-{\bf
A^2})+\frac{f(\gamma)f(\gamma-1)}{2\gamma+1}((\gamma+1)
r(\gamma)+{\bf A^2})
\end{equation}
Our goal is to make $T=1$. Having expressions for $r(\gamma)$ and
$\bf A^2$
\begin{eqnarray}
&r(\gamma)=2\gamma(\gamma+1)-h\nonumber\\
&{\bf
A^2}=\alpha^2+h(\gamma(\gamma+1)+1)-\gamma^2(\gamma+1)^2-2\gamma(\gamma+1),
\end{eqnarray}
we can rewrite the equation $T=1$ in the following form:
\begin{eqnarray}\label{40}
f(\gamma)f(\gamma+1)[\alpha^2+h(\gamma+1)^2-(\gamma+1)^2((\gamma+1)^2-1)]\nonumber\\-
f(\gamma)f(\gamma-1)[\alpha^2+h\gamma^2-\gamma^2(\gamma^2-1)]=-(2\gamma+1).
\end{eqnarray}
The first and the second terms in the L.H.S. of (\ref{40}) differ by
shift of $\gamma$ by 1, therefore we immediately obtain
\begin{equation}\label{41}
f(\gamma)f(\gamma-1)=\frac{\mu-\gamma^2}{\alpha^2+h\gamma^2-\gamma^2(\gamma^2-1)},
\end{equation}
where $\mu$ does not depend of $\gamma$. Before solving this
equation for $f(\gamma)$ let us find the square of operator $\bf R$
\begin{eqnarray}
{\bf R^2}=f^{1/2}(\gamma)A_i f(\gamma)A_if^{1/2}(\gamma).
\end{eqnarray}
Using again equation (\ref{33})we can move $A_i$ to the right
through $f(\gamma)$ and after some algebraic transformation we
arrive at
\begin{eqnarray}
{\bf R^2}=\frac{f(\gamma)f(\gamma+1)}{2\gamma+1}(\gamma+1)[{\bf
A^2}-\gamma (\gamma)]\nonumber\\
+\frac{f(\gamma)f(\gamma-1)}{2\gamma+1}\gamma[{\bf A^2}+(\gamma+1)
r(\gamma)]
\end{eqnarray}
Here we can use the result (\ref{41}) and complete this calculation
\begin{equation}
{\bf R^2}=\mu-1-\gamma(\gamma+1),
\end{equation}
therefore we get
\begin{equation}
{\bf R^2+L^2}=\mu-1
\end{equation}
the result which was expected. Now it is time to choose $\mu$. The
equation (\ref{41}) could be written as follows
\begin{eqnarray}
f(\gamma)f(\gamma-1)=\frac{\mu-\gamma^2}{(k_1-\gamma^2)(\gamma^2-k_2)},\nonumber\\
k_{1,2}=\frac{h+1}{2}\pm \sqrt{\frac{(h+1)^2}{4}+\alpha^2}.
\end{eqnarray}
This formula wakes the reminiscence of the classical case. The
choice
\begin{equation}\label{47}
\mu=\frac{h+1}{2}+ \sqrt{\frac{(h+1)^2}{4}+\alpha^2}
\end{equation}
gives us
\begin{equation}
{\bf R^2+L^2}=-1+\frac{h+1}{2}+ \sqrt{\frac{(h+1)^2}{4}+\alpha^2}
\end{equation}
and as a result, the expression for Hamiltonian via ${\bf R^2+L^2}$:
\begin{equation}
h={\bf R^2+L^2}-\frac{\alpha^2}{{\bf R^2+L^2}+1}.
\end{equation}
Now let us discuss this representation of Hamiltonian. The operators
$\bf R$ and $\bf L$ forms the algebra $SO(4)$, which is the direct
sum of two algebras $SO(3)$, so we can introduce instead of $\bf R$
and $\bf L$ another pair of operators $\bf M,N$
\begin{equation}
{\bf M}=\frac{1}{2}{\bf(L+R)},\qquad {\bf N}=\frac{1}{2}{\bf(L-R)}
\end{equation}
such that
\begin{eqnarray}
&[M_i,M_j]=i\epsilon_{ijk}M_k,\qquad[N_i,N_j]=i\epsilon_{ijk}N_k \nonumber\\
 &\quad [M_i,N_j]=0.
\end{eqnarray}
The two Casimirs of $S0(4)$ in terms of $\bf M,N$ are
\begin{equation}
C_1={\bf M^2},\qquad C_2={\bf N^2}
\end{equation}
and because $\bf RL=0$,
\begin{equation}
C_1=C_2=\frac{1}{4}({\bf R^2+L^2}).
\end{equation}
The spectrum of $C_1 (C_2)$ is given by
$k(k+1),k=0,\frac{1}{2},1...$. The representation, characterized by
$k$ contains all angular momenta $l=0,1,...2k$ with multiplicity 1.
As a result, the spectrum of Hamiltonian $h$ will be given by
\begin{equation}\label{54}
h=4k(k+1)-\frac{\alpha^2}{4k(k+1)+1}=2k(2k+2)-\frac{\alpha^2}{(2k+1)^2}.
\end{equation}
Introducing another quantum number $n=2k+1, n=1,2...$, we can
present (\ref{54}) in the form
\begin{equation}
h=(n-1)(n+1)-\frac{\alpha^2}{n^2}
\end{equation}
which coincides with Sch\"{o}dinger's result \cite{Schrod}.

The last thing which has to be mentioned here if the explicit form
of function $f(\gamma)$, although we can easily avoid its
construction. Using equation (\ref{33}) we can find vector $\bf R$,
knowing only bilinear combinations like $f(\gamma)f(\gamma\pm 1)$.
But for completeness we shall produce the result especially because
it worth to be mentioned. With our choice of $\mu$ ---(\ref{47}),
the equation (\ref{41}) takes the following form:
\begin{eqnarray}
f(\gamma)f(\gamma-1)=\frac{1}{(\gamma^2+
\sqrt{\frac{(h+1)^2}{4}+\alpha^2}-\frac{h+1}{2})}.
\end{eqnarray}
The expression $\sqrt{\frac{(h+1)^2}{4}+\alpha^2}-\frac{h+1}{2})$ is
always positive, so denoting  it as $\rho^2$,we have
\begin{equation}
f(\gamma)f(\gamma-1)=\frac{1}{\gamma^2+\rho^2}
\end{equation}
The solution of this equation is given by
\begin{equation}
f(x)=
\displaystyle\frac{i}{x-i\rho}\frac{\Gamma(\frac{x-i\rho+1}{2})}
{\Gamma(\frac{x-i\rho}{2})}
\frac{\Gamma(-\frac{x+i\rho}{2})}{\Gamma(-\frac{x+i\rho-1}{2})}.
\end{equation}
Proof could be done by direct substitution into equation.

\section*{Acknowledgments}
The author is grateful to professors A.K.Likhoded and A.V.Razumov
for their comments and fruitful discussions, professor M.Santander
for drawing my attention to this problem and providing with
references. This work  was supported by the programme ENTER-2004/04EP-48,
 E.U.-European Social Fund(75{\%}) and Greek Ministry of development-GSRT (25{\%})
and by RFFI grant 07-01-00234.

\appendix
\section{Appendix}

Let us consider a function $F({\bf L^2})$ and multiply it from the
left by an operator $\bf A$ which is a vector with respect to $\bf
L$ and satisfies the condition $\bf AL=0$. In general the following
relation exists:
\begin{equation}\label{A1}
A_i F({\bf L^2})=S({\bf L^2})A_i+T({\bf L^2})i\epsilon_{ijk}L_j A_k
\end{equation}
where the functions $S,T$ is defined by $F$. The limitation to the
case of $\bf AL=0$ is not essential, but as we do not need this
general case for our purpose. Let the function $F({\bf L^2})$ be
represented in the form
\begin{equation}\label{A2}
F({\bf L^2})=\int d\alpha \phi(\alpha)e^{ix\alpha {\bf L^2}}
\end{equation}
Thus, in order to derive (\ref{A1}) we need to consider only the
case of exponent of $\bf L^2$. Multiplying the exponent by $\bf A$
from the left, we have
\begin{equation}\label{A3}
A_i e^{i\alpha {\bf L^2}}=f(\alpha,{\bf L^2})A_i+g(\alpha,{\bf
L^2})i \epsilon_{ijk}L_j A_k ,
\end{equation}
where the functions $f,g$ have to be defined. Now let us
differentiate both sides of (\ref{A3}) over $\alpha$
\begin{eqnarray}\label{A4}
&\partial_{\alpha}A_i e^{i\alpha {\bf L^2}}=iA_i e^{i\alpha {\bf
L^2}}{\L^2}=i\left[f(\alpha,{\bf L^2})A_i+g(\alpha,{\bf L^2})i
\epsilon_{ijk}L_j A_k\right]{\bf L^2}\nonumber\\
&=i\left[f(\alpha,{\bf L^2})({\bf L^2}+2)+2g(\alpha,{\bf L^2}){\bf
L^2}\right]A_i\nonumber\\
&+i\left[2f(\alpha,{\bf L^2}) +g(\alpha,{\bf L^2})\right]i
\epsilon_{ijk}L_j A_k\nonumber\\
& =\partial_{\alpha}f(\alpha,{\bf
L^2})A_i+\partial_{\alpha}g(\alpha,{\bf L^2})i \epsilon_{ijk}L_j
A_k,
\end{eqnarray}
where we have used the equation
\begin{equation}\label{A5}
A_i{\bf L^2}=({\bf L^2}+2)A_i+2i\epsilon_{ijk}L_j A_k
\end{equation}
From (\ref{A4}) follow two equation for functions $f,g$:
\begin{eqnarray}\label{A6}
&\partial_{\alpha}f(\alpha,{\bf L^2})=i\left[f(\alpha,{\bf
L^2})({\bf
L^2}+2)+2g(\alpha,{\bf L^2}){\bf L^2}\right]\nonumber\\
&\partial_{\alpha}g(\alpha,{\bf L^2})=i\left[2f(\alpha,{\bf L^2})
+g(\alpha,{\bf L^2})\right],\nonumber\\
&f(0,{\bf L^2})=1, \qquad g(0,{\bf L^2})=0
\end{eqnarray}
Further it will be convenient to introduce instead of $\bf L^2$ the
operator $\gamma$
\begin{equation}\label{A7}
\gamma=\sqrt{({\bf L^2}+\frac{1}{4})}-\frac{1}{2},\qquad {\bf
L^2}=\gamma(\gamma+1).
\end{equation}
In terms of $\gamma$ the solution of (\ref{A6}) has the following
form:
\begin{eqnarray}\label{A8}
&f(\alpha,\gamma(\gamma+1))=\frac{1}{2\gamma+1}
\left[(\gamma+1)e^{i\alpha(\gamma+1)(\gamma+2)}+
\gamma e^{i\alpha\gamma(\gamma-1)}\right]\nonumber\\
&g(\alpha,\gamma(\gamma+1))=\frac{1}{2\gamma+1}\left[e^{i\alpha(\gamma+1)(\gamma+2)}-
e^{i\alpha\gamma(\gamma-1)}\right].
\end{eqnarray}
Substituting (\ref{A8}) into (\ref{A3}) we receive:
\begin{eqnarray}
A_i
e^{i\alpha\gamma(\gamma+1)}=e^{i\alpha(\gamma+1)(\gamma+2)}\frac{1}{2\gamma+1}
\left[(\gamma+1)A_i+i\epsilon_{ijk}L_j A_k\right]\nonumber\\
+e^{i\alpha\gamma(\gamma-1)}\frac{1}{2\gamma+1} \left[\gamma
A_i-i\epsilon_{ijk}L_j A_k\right].
\end{eqnarray}
Thus, moving the vector operator through the exponent of
$\gamma(\gamma+1)$ produces two term, one with $\gamma$ shifted by
$+1$, the other by $-1$. Would we had the operator $\bf A$ such that
$\bf AL\not=0$ the third term will appear , where $\gamma$ will not
be shifted. Making the Fourier transformation we extend the result
for arbitrary function
\begin{eqnarray}
A_i F(\gamma)=F(\gamma+1)\frac{1}{2\gamma+1}
\left[(\gamma+1)A_i+i\epsilon_{ijk}L_j A_k\right]\nonumber\\
+F(\gamma-1)\frac{1}{2\gamma+1} \left[\gamma A_i-i\epsilon_{ijk}L_j
A_k\right].
\end{eqnarray}
The analogous formula exists also for right multiplication.

\end{document}